\documentclass[preprint,showpacs,preprintnumbers,amsmath,amssymb]{revtex4}
\usepackage{amsmath,amssymb,graphics,epsfig,subfigure}

\begin{document}
\newcommand {\nn}    {\nonumber}
\renewcommand{\baselinestretch}{1.3}

\title{Localization of Gravity and Bulk Matters on a Thick Anti-de Sitter Brane}

\author{ Yu-Xiao Liu\footnote{liuyx@lzu.edu.cn},
         Heng Guo\footnote{guoh2009@lzu.cn,Corresponding author},
         Chun-E Fu\footnote{fuche08@lzu.cn},
         Hai-Tao Li\footnote{liht07@lzu.cn}}.\\
        \affiliation{Institute of Theoretical Physics,
              Lanzhou University, Lanzhou 730000,
             People's Republic of China}


\begin{abstract}

In this paper, we investigate the localization and the mass spectra
of gravity and various bulk matter fields on a thick anti-de Sitter
(AdS) brane, by presenting the mass-independent potentials of the
Kaluza-Klein (KK) modes in the corresponding Schr\"{o}dinger
equations. For gravity, the potential of the KK modes tends to
infinity at the boundaries of the extra dimension, which leads to an
infinite number of the bound KK modes. Although the gravity zero
mode cannot be localized on the AdS brane, the massive modes are
trapped on the brane. The scalar perturbations of the thick AdS
brane have been analyzed, and the brane is stable under the scalar
perturbations. For spin-0 scalar fields and spin-1 vector fields,
the potentials of the KK modes also tend to infinity at the
boundaries of the extra dimension, and the characteristic of the
localization is the same as the case of gravity. For spin-1/2
fermions, by introducing the usual Yukawa coupling
$\eta\bar{\Psi}\phi\Psi$ with the positive coupling constant $\eta$,
the four-dimensional massless left-chiral fermion and massive Dirac
fermions are obtained on the AdS thick brane.
\end{abstract}




\maketitle

\section{Introduction}

The idea that our observed four-dimensional Universe might be a
3-brane, embedded in a higher dimensional space-time (the bulk),
provides new insights into solving the gauge hierarchy and
cosmological constant
problems~\cite{RubakovPLB1983136,VisserPLB1985,ADD,
Randjbar-DaemiPLB1986,rs,Lykken,AntoniadisPLB1990,CosmConst}. In the
Randall-Sundrum (RS) brane world model, the zero mode of gravity is
localized on the brane, which reproduces the standard Newtonian
gravity on the brane~\cite{rs}. But in this model, the brane is very
ideal because its thickness is neglected. In the most fundamental
theory, there seems to exist a minimum scale of length, thus the
thickness of a brane should be considered in more realistic field
models. For this reason, more natural thick brane scenarios have
been investigated~\cite{De_Wolfe_PRD_2000,PRD_Stojkovic,
Gremm_2000,paper2002,PRD_koyama,Csaki_NPB_2000,
0005016,PRL845928,CamposPRL2002,Wang_PRD,varios,ThickBrane,Bazeia,
Volkas0705.1584,GherghettaPRL2000,Neupane,ThickBraneWeyl,Cvetic,
PRD0709.3552,ThickBrane4,KobayashiPRD2002,RandallJHEP2001,0910.0363,
liu_0911.0269,Brandhuber_JHEP,Diaz_1999,Oda_PRD,
Liu0907.1952,1004.0150,zhong_fRBrane,Singleton,Guo_BentBrane,1009.1684}.
For some comprehensive reviews about thick branes, see Refs.
\cite{0812.1092,0904.1775,0907.3074,1003.1698,brane_book,1004.3962}.

In brane world theory, only gravity is free to propagate in both the
brane and bulk space-time, however, all the matter fields
(electromagnetic, Yang-Mills, etc.) in our four-dimensional Universe
are confined to the 3-brane with no contradiction with present
gravitational experiments. Hence, in order to build up the standard
model, various bulk matter fields should be localized on branes by a
natural mechanism. Generally, the massless scalar fields
\cite{BajcPLB2000} could be trapped on branes of different types.
Spin-1 Abelian vector fields can be localized on the RS brane in
some higher dimensional cases \cite{OdaPLB2000113} or on the thick
de Sitter brane and Weyl thick brane \cite{Liu0708}. It is important
to study the localization problem of the spin-1/2 fermions. Without
introducing the scalar-fermion coupling, fermions cannot be
localized on branes in five and six dimensions
\cite{BajcPLB2000,OdaPLB2000113,NonLocalizedFermion,
IchinosePRD2002,Ringeval,
RandjbarPLB2000,KoleyCQG2005,Liu0708,DubovskyPRD2000,0803.1458,
LiuJHEP2007,0901.3543,Liu0907.0910,Koley2009,0812.2638,
LiuJCAP2009,Liu0803,Singleton_fermion,JackiwPRD1976,liu_0909.2312,Guo_jhep,Liu_2010,
zhao_CQG,Fu_KR}. In some cases, there may exist a single bound state
and a continuous gapless spectrum of massive fermion Kaluza-Klein
(KK) modes \cite{ThickBraneWeyl,Liu0708}. In some other cases, one
can obtain finite discrete KK modes (mass gap) and a continuous
gapless spectrum starting at a positive $m^{2}$
\cite{ThickBrane4,LiuJCAP2009,Liu0803}.

Most investigations mentioned above mainly focused on flat branes.
In Ref. \cite{Wang_PRD}, the solutions of de Sitter (dS) and anti-de
Sitter (AdS) $3$-branes were presented, and the localization of
gravity on the dS $3$-brane was also studied. In this paper, in
order to show the rich structures of the AdS $3$-brane from the
other points of view, we would like to investigate the localization
problem of gravity and various spin matter fields (scalars, vectors
and fermions) on the brane. For this AdS $3$-brane solution, the
behavior of the warp factor is related with a parameter $\delta$.
For $\delta>1$, the warp factor is divergent at $z=\pm z_b$ the
boundaries of the extra dimension, and the energy density has a lump
at $z\approx0$; i.e., the brane is located at $z\approx0$. For
$\delta<0$, the warp factor tends to zero far away from the brane,
but the energy density has no lump at $z\approx0$ and diverges at
$z=\pm z_b$, so this configuration cannot be considered as a thick
brane. Hence, in this paper, we only consider that the parameter is
constrained as $\delta>1$. For gravity, the zero mode is not
localized on the thick AdS brane; however, the massive KK modes can
be localized on the brane, and the mass spectrum consists of an
infinite number of discrete bound states. For free scalar fields and
vector fields, all the KK modes are bound states and the massive
modes, which could be trapped on the brane. For spin-1/2 fermions
coupling with the background scalar by the usual Yukawa coupling
$\eta\bar{\Psi}\phi\Psi$, the left-chiral fermion massless mode can
be localized on the brane, and both left- and right-chiral fermions
are also localized on the brane, which can constitute the
four-dimensional massive Dirac fermions.

The organization of this paper is as follows: In Sec.
\ref{SecModel}, we first review the thick AdS $3$-brane in
five-dimensional space-time. Then, in Sec.
\ref{LocalizationGravity}, we investigate the localization and the
mass spectra of gravity on the thick AdS $3$-brane. In Sec.
\ref{SecLocalization}, the localization and the mass spectra of
various bulk matter fields are investigated. For scalars and
vectors, we give the analytical expressions for the KK modes and the
mass spectrums. For fermions, the analytical formulations of the
massless modes are also obtained. Finally, our conclusion is given
in Sec. \ref{SecConclusion}.

\section{Review of the thick anti-de Sitter 3-brane}
\label{SecModel}

We start with the following five-dimensional action of thick branes,
which are generated by a real scalar field $\phi$ with a scalar
potential $V(\phi)$,
\begin{equation}
 S= \int d^5 x \sqrt{-g}\left[\frac{1}{2\kappa_5^2} R-\frac{1}{2}
     g^{MN}\partial_M \phi \partial_N \phi - V(\phi) \right],
\label{action}
\end{equation}
where $R$ is the five-dimensional scalar curvature and $\kappa_5^2=8
\pi G_5$ with $G_5$ the five-dimensional Newton constant. The most
general five-dimensional metric compatible with an AdS$_{4}$
symmetry can be taken as
\begin{eqnarray}\label{linee}
ds^{2}&=& g_{MN}dx^{M}dx^{N}
      = \text{e}^{2A(z)}[\hat{g}_{\mu\nu}(x)dx^\mu dx^\nu + dz^2]
      \nonumber \\
      &=& \text{e}^{2A(z)}\left[ \text{e}^{2H x_{3}}
             (-dt^{2}+dx_{1}^{2}+dx_{2}^{2}) +dx_{3}^{2}
             +dz^{2}\right],~~~
\end{eqnarray}
where $\text{e}^{2A(z)}$ is the warp factor, $H$ is the AdS$_{4}$
parameter and $z$ stands for the extra coordinate. We suppose that
the scalar field is a function of $z$ only, i.e., $\phi=\phi(z)$.
The four-dimensional cosmology constant can be expressed as
$\Lambda_{4}=-3H^{2}$. By considering the action (\ref{action}) and
the metric (\ref{linee}), the filed equations reduce to the
following coupled nonlinear differential equations:
\begin{subequations}\label{EinsteinEq}
\begin{eqnarray}
\label{EinsteinEq_a}
\phi'^{2}&=&\frac{3}{\kappa_{5}^{2}}(A'^{2}-A''+H^{2}),\\
\label{EinsteinEq_b}
 V(\phi)&=& -\frac{3\text{e}^{-2A}}{2\kappa_{5}^{2}}
            (3H^{2}+3A'^{2}+A''),\\
\label{EinsteinEq_c}
 \frac{d V(\phi)}{d \phi}&=&\text{e}^{-2A}(3A'\phi'+\phi''),
\end{eqnarray}
\end{subequations}
where the prime denotes the derivative with respect to $z$. These
equations are not independent. The relationship
between these equations was discussed in \cite{Csaki_NPB_2000}.

Here we set $\kappa_{5}^{2}=1$. A thick AdS brane solution in
five-dimensional space-time for the potential
\begin{eqnarray}\label{Vphi}
V(\phi)=-\frac{3(1+3\delta)H^{2}}{2\delta}\cosh^{2(1-\delta)}
            \left(\frac{\phi}{\phi_{0}} \right)
\end{eqnarray}
was found in Ref. \cite{Wang_PRD}:
\begin{eqnarray}
\label{warpfactor}
A(z)&=& -\delta\ln\left|\cos \bar{z} \right|,
\\
\label{scalarsolution}
\phi(z)&=&\phi_{0}\text{arcsinh}(\tan\bar{z}),
\end{eqnarray}
where
\begin{eqnarray}
\phi_{0}&\equiv&\sqrt{3\delta(\delta-1)}, \\
\bar{z}&\equiv& \frac{H}{\delta}z,
\end{eqnarray}
and the parameter $\delta$ satisfies $\delta >1$ or $\delta <0$. The
range of the extra dimension is $-z_{b}\leq z\leq z_{b}$ with
$z_{b}=\left|\frac{\delta\pi}{2H}\right|$. The metric with this
choice of $A(z)$ has a naked singularity at $\pm z_{b}$. This
singularity is very similar to the one in Ref.~\cite{Gremm_2000} and
the one encountered in the AdS flow to $N=1$ super Yang-Mills
theory~\cite{NPB569}. Gremm supported that this might indicate that
it can be resolved either by lifting the five-dimensional geometry
to ten dimensions or by string theory~\cite{Gremm_2000}. From
(\ref{scalarsolution}), it can be seen that the background scalar
field diverges at the boundaries of the extra dimension $\pm z_{b}$.
In Ref.~\cite{Gremm_2000}, this divergence can indicate that the
compactification manifold shrinks to zero or becomes infinitely
large, so that the five-dimensional truncation comes to be invalid.
There are some examples that singularities in five dimensions
actually correspond to nonsingular ten-dimensional
geometries~\cite{TenDimension}.

As is well-known, the energy-momentum tensor for a scalar field is
energetically equivalent to an anisotropic fluid,
$T_{MN}=-\rho(g_{MN}+z_{M}z_{N})+pz_{M}z_{N}$, where
$z_{M}=\text{e}^{A}\delta_{M}^{5}$ and
\begin{eqnarray}
\label{EnergyDensity}
 \rho&\equiv&\frac{1}{2}\big[\text{e}^{-2A}\phi'^{2}+2V(\phi)\big] \nonumber \\
             &=& -\frac{3H^{2}(1+\delta)}{\delta}\cos^{2(\delta-1)}(\bar{z}),   \\
    p&\equiv&\frac{1}{2}\big[\text{e}^{-2A}\phi'^{2}-2V(\phi)\big] \nonumber \\
             &=& 6H^{2}\cos^{2(\delta-1)}(\bar{z}).
\end{eqnarray}
From the above expressions, it can be shown that, for $\delta<0$ or
$\delta>1$, the corresponding energy-momentum tensor satisfies the
weaker energy condition $T_{MN}\zeta^{M}\zeta^{N}\geq0$ with
$\zeta^{M}$ an arbitrary null vector. For $-1<\delta<0$, the weak
energy condition $T_{MN}\xi^{M}\xi^{N}\geq0$ is satisfied, where
$\xi^{M}$ is an arbitrary future-directed timelike or null vector,
and for $ -\frac{1}{3}<\delta<0$ the dominant energy condition
$T_{MN}\xi^{M}\eta^{N}\geq0$ with $\eta^{M}$ an arbitrary
future-directed timelike or null vector is satisfied. It is clear
that, for $\delta>1$, the energy density $\rho$ has a lump at
$|z|\approx0$ and tends to zero at the boundaries of the extra
dimension $z=\pm z_{b}$, which shows that the thick 3-brane locates
at $|z|\approx0$. However, for $\delta<0$, the energy density has no
lump at $z\approx0$ and diverges at $\pm z_{b}$, and it is difficult
to regard this configuration as a thick brane, so in the following
sections, we only consider the case of $\delta>1$.

\section{Localization of gravity on the thick anti-de Sitter brane}
\label{LocalizationGravity}

In this section, the localization of gravity will be investigated by
presenting the mass-independent potential of the KK modes for
gravitons in the corresponding Schr\"{o}dinger equation.

\begin{figure}[htb]
\begin{center}
\subfigure[]{\label{fig_V0_3}
\includegraphics[width=7cm]{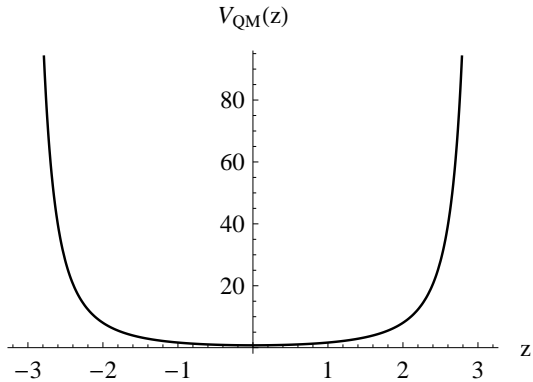}}
\end{center}\vskip -5mm
\caption{The shape of the potential of the gravity KK modes
$V_{\text{QM}}(z)$. The parameters are set to $H=1$ and $\delta=2$.}
 \label{fig_Scalar_V0Chi0}
\end{figure}

\begin{figure*}[htb]
\begin{center}
\subfigure[$n=1$]{\label{fig_chi0a}
\includegraphics[width=3.5cm]{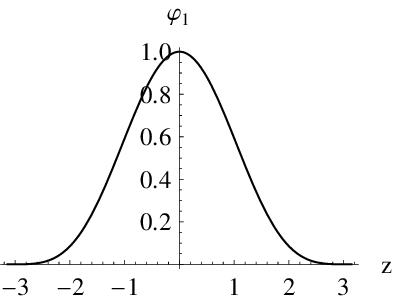}}
\subfigure[$n=2$]{\label{fig_chi1a}
\includegraphics[width=3.5cm]{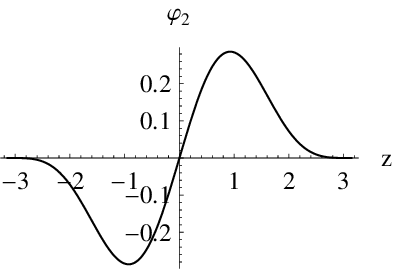}}
\subfigure[$n=3$]{\label{fig_chi2a}
\includegraphics[width=3.5cm]{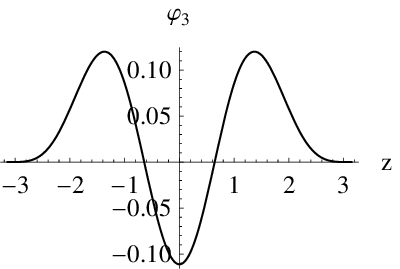}}
\subfigure[$n=4$]{\label{fig_chi3a}
\includegraphics[width=3.5cm]{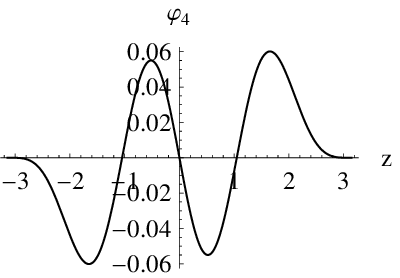}}
\end{center}\vskip -5mm
\caption{The shapes of the KK modes of the graviton
$\varphi_{n}(z)$. The parameters are set to $H=1$ and $\delta=2$.}
 \label{fig_china}
\end{figure*}

Let us consider the following perturbed metric:
\begin{eqnarray}
ds^{2}=(g_{MN}+\delta g_{MN})dx^{M}dx^{N}.
\end{eqnarray}
As Ref.~\cite{rs}, we impose the axial gauge constraint
 $\delta g_{5M}=0$, and then write the total metric in the
form
\begin{eqnarray}
ds^{2}=\text{e}^{2A(z)}[(\hat{g}_{\mu\nu}+h_{\mu\nu})dx^{\mu}dx^{\nu}+dz^{2}],
\end{eqnarray}
where $h_{\mu\nu}$ denotes the metric perturbation. Then, under the
transverse-traceless gauge condition
$h^{\mu}_{~\mu}=\partial^{\mu}h_{\mu\nu}=0$, the equation for the
perturbation $h_{\mu\nu}$ takes the following form~\cite{PRD_koyama}
\begin{eqnarray}\label{EqGravity}
 [\partial_{z}^{2}+3A'(z)\partial_{z} +\Box
              +2H^{2}]h_{\mu\nu}(x^{\alpha},z)=0,
\end{eqnarray}
where $\Box\equiv\hat{g}^{\mu\nu}\nabla_{\mu}\nabla_{\nu}$ and
$\nabla_{\mu}$ denotes the covariant derivative with respect to the
four-dimensional metric $\hat{g}_{\mu\nu}$. By making use of the KK
decomposition
\begin{eqnarray}
 h_{\mu\nu}(x^{\alpha},z)=\text{e}^{-\frac{3}{2}A(z)}
                       \epsilon_{\mu\nu}(x^{\alpha})\varphi(z)
\end{eqnarray}
with $\epsilon_{\mu\nu}(x^{\alpha})$ satisfying the
transverse-traceless condition, from Eq. (\ref{EqGravity}) we can
get the following four-dimensional equation
\begin{eqnarray}
 \Box \epsilon_{\mu\nu}(x^{\alpha})
       +2H^{2}\epsilon_{\mu\nu}(x^{\alpha})
  =m^{2}\epsilon_{\mu\nu}(x^{\alpha}),
\end{eqnarray}
where $m$ is the mass of the KK modes. The Schr\"{o}dinger-like
equation for the fifth-dimensional sector is also obtained
\begin{eqnarray}\label{GravitySchEq}
 (-\partial_{z}^{2}+V_{\text{QM}})\varphi(z)=m^{2}\varphi(z)
\end{eqnarray}
with the potential
\begin{eqnarray}\label{VQM}
 V_{\text{QM}}&=&\frac{3}{2}A''(z)+\frac{9}{4}A'^{2}(z) \nonumber \\
              &=&\frac{3H^{2}}{4\delta}
                \left[(2+3\delta)\sec^{2}\bar{z}-3\delta\right].
\end{eqnarray}
The shapes of the potential for the KK modes are plotted in
Fig.~\ref{fig_Scalar_V0Chi0}. The potential $V_{\text{QM}}(z)$ has
minimum $\frac{3H^{2}}{2\delta}$ (positive value) at the location of
the brane ($z=0$), and tends to positive infinity at the boundaries
of the fifth dimension ($z=\pm z_{b}$), so the potential has no
negative value at the location of the brane, and it cannot trap the
zero mode of gravity, but it can trap the massive modes. Eq.
(\ref{GravitySchEq}) can be turned into the following equation with
$E_{n}^{2}=m_{n}^{2}+\frac{9H^{2}}{4}$:
\begin{eqnarray}
\label{SchEqScalar2}
\left[-\partial_{z}^{2}+\frac{3H^{2}}{4\delta}(2+3\delta)\sec^{2}\bar{z}\right]
                       \varphi_{n}(z)=E_{n}^{2}\varphi_{n}(z).
\end{eqnarray}
When the parameter $\delta$ satisfies $\delta>1$, the solutions of
the gravity KK modes are found to be
\begin{eqnarray}
\varphi_{n}(z)&\propto&
               ~_{2}\text{F}_{1}\Big[1-n, 1+n+3\delta, \frac{3(1+\delta)}{2},
               \frac{1-\sin\bar{z}}{2}\Big] \nonumber\\
              &&~\times
                 \cos^{1+\frac{3\delta}{2}} (\bar{z}),
\end{eqnarray}
where $n=1,2,\cdots~$.
Then the mass spectrum of bound states is found to be
\begin{eqnarray}
E_{n}=\frac{H}{\delta} \left(n+\frac{3}{2}\delta\right),
\end{eqnarray}
or
\begin{eqnarray}
m_{n}=\frac{H}{\delta}\sqrt{n(n+3\delta)}.
\end{eqnarray}
It can be found that the ground state is the massive mode with
$m_{1}=\frac{H}{\delta}\sqrt{1+3\delta}$. All the KK modes are bound
states, and localized on the thick AdS brane. The mass spectrum of
the KK modes is discrete. The shapes of the gravity KK modes with lower mass are
shown in Fig.~\ref{fig_china}.

From the above discussion, $m^{2}$ is always positive, and following
the arguments given in Ref.~\cite{PRD_koyama}, it can be shown that
in the present case all the corresponding perturbation modes are
stable. Now we need to further investigate the stability of this AdS
brane, i.e., the scalar perturbation of the systems should be
examined. The perturbed metric is given by \cite{PRD_koyama}
\begin{eqnarray}
 ds^{2}=
 \text{e}^{2A}[(1+2\alpha)dz^{2}+(1+2\beta)\hat{g}_{\mu\nu}dx^{\mu}dx^{\nu}].
\end{eqnarray}
As shown in Ref.~\cite{PRD_koyama}, the following corresponding
linearized five-dimensional Einstein-scalar equations can be
obtained:
\begin{eqnarray}
 &~& \delta\phi=\frac{3}{\phi'}(\alpha A'-\beta'), \quad\quad
 \alpha+2\beta=0, \\
 &~&  \beta'' + \Box\beta
        -\left(3A'+2\frac{\phi''}{\phi'}\right)\beta' \nonumber\\
     && +\left(4A'\frac{\phi''}{\phi'}-4A''-6H^{2}\right)\beta=0,
 \label{masterEq}
\end{eqnarray}
where $\delta\phi$ denotes perturbations of the background scalar
field. In order to examine the stability of this system, Eq.
(\ref{masterEq}) is transformed into a form of the Schr\"{o}dinger
equation:
\begin{eqnarray}\label{SchEqDeltaPhi}
 [-\partial_{z}^{2}+V_{\text{eff}}(z)]\omega(x^{\mu},z)
  =\Box \omega(x^{\mu},z),
\end{eqnarray}
where $\omega(x^{\mu},z)$ and $V_{\text{eff}}(z)$ are defined as
\begin{eqnarray}
\label{omega}
 \omega &\equiv&
 \frac{1}{\phi'(z)}\text{e}^{\frac{3A}{2}}\beta(x^{\mu},z),
   \\
\label{Veff}
 V_{\text{eff}}&\equiv&
  -\frac{5}{2}A''+\frac{9}{4}A'^{2}+A'\frac{\phi''}{\phi'}
  -\frac{\phi'''}{\phi'}+2\left(\frac{\phi''}{\phi'} \right)^{2}
   + 6H^{2} ~~~\nonumber\\
   &=& \frac{H^{2}}{4\delta^{2}}\left[(15\delta^{2}-4\delta-4)
       + 3\delta(3\delta-2)\sec^{2}\bar{z}\right].
\end{eqnarray}
From above expression (\ref{Veff}), it is clear that because the
parameter $\delta$ satisfies $\delta>1$, the effective potential
$V_{\text{eff}}(z\rightarrow\pm z_{b})\rightarrow +\infty$, and has
minimum $\frac{H^{2}}{2\delta^{2}}(3\delta-2)(4\delta+1)$ (positive
value) at $z=0$.

We decompose $\omega(x^{\mu},z)$ in Eq. (\ref{SchEqDeltaPhi}) as
\begin{eqnarray}
 \omega(x^{\mu},z)=f(z)X(x^{\mu}),
\end{eqnarray}
and the following equations can be gotten
\begin{eqnarray}
 \label{4dKGScalar}
 \Box X(x^{\mu})&=&m^{2}X(x^{\mu}),\\
 \label{SchEqBackSca}
 -f''(z)+V_{\text{eff}}(z)f(z)&=& m^{2}f(z)
\end{eqnarray}
with $m$ the four-dimensional mass. It is known that Eq.
(\ref{4dKGScalar}) can be solved with suitable harmonic functions.
There is the Breitenlohner-Freedman bound which allows the tachyonic
mass to some extent from the condition of the normalization
\cite{PRD_koyama,BFbound}. The mass $m$ is bounded as
\begin{eqnarray}
 m^{2}\geq -\frac{9}{4},
\end{eqnarray}
which means that even when there are solutions with
$-\frac{9}{4}\leq m^{2}<0$, such solutions are stable in spite of
the tachyonic mass.

By substituting the effective potential~(\ref{Veff}) into Eq.
(\ref{SchEqBackSca}), we can get
\begin{eqnarray}
 \label{SchEqBackScaI}
 \left[-\partial_{z}^{2}+
  \frac{3H^{2}}{4\delta}(3\delta-2)\sec^{2}\bar{z}\right]
   f_{n}(z)&=& E_{n}^{2}f_{n}(z),~~~
\end{eqnarray}
with
$E_{n}^{2}=m_{n}^{2}-\frac{H^{2}}{4\delta^{2}}(15\delta^{2}-4\delta-4)$.
Since $\delta>1$, the solutions of the above Schr\"{o}dinger Eq.
(\ref{SchEqBackScaI}) can be expressed as
\begin{eqnarray}
 f_{n}(z)\propto \cos^{\frac{3\delta}{2}}(\bar{z})~
             _{2}\text{F}_{1}
             \left[ -n, n+3\delta, \frac{1+3\delta}{2},
                     \frac{1-\sin\bar{z}}{2} \right],
                    ~
\end{eqnarray}
where $n=0,1,2,\cdots~$.
It is known that the corresponding perturbation modes tend to zero
at the boundaries of the extra dimension, i.e., they are bound
states, and all the scalar modes of the metric fluctuations can be
localized on the AdS brane. The mass spectrum of the bound states is
found to be
\begin{eqnarray}
 E_{n}=\frac{H}{\delta}\left(n+\frac{3\delta}{2}\right),
\end{eqnarray}
or
\begin{eqnarray}
 m_{n}=\frac{H}{\delta}
        \sqrt{n^{2}+6\delta^{2}+(3n-1)\delta-1}.
\end{eqnarray}
It is clear that the mass of the ground state is
 $m_{0}=\frac{H}{\delta}\sqrt{6\delta^{2}-\delta-1}$,
and
 $m_{n}^{2}\geq \frac{H^{2}}{\delta^{2}}[n^{2}+6\delta^{2}+(3n-1)\delta-1]>0$,
so it is shown that the thick AdS brane is stable under the
scalar perturbations.

\section{Localization of various matters on the thick anti-de Sitter brane}
\label{SecLocalization}

In this section, we will investigate the character of the
localization of the various bulk matter fields on the thick AdS
brane. Spin-0 scalars, spin-1 vectors and spin-1/2 fermions will be
considered by means of the gravitational interaction. Certainly, we
have implicitly assumed that the various bulk matter fields
considered below make little contribution to the bulk energy so that
the solutions given in the previous section remain valid even in the
presence of bulk matter fields. The mass spectra of the various
matter fields on the thick brane will also be discussed by
presenting the potential of the corresponding Schr\"{o}dinger
equation for the KK modes of the various matter fields.

\subsection{Spin-0 scalar fields}

We first consider the localization of real scalar fields on the
thick brane obtained in the previous section, then turn to vectors
and fermions in the next subsections. Let us start by considering
the action of a massless real scalar coupled to gravity:
\begin{eqnarray}\label{action_scalar}
S_{0}=-\frac{1}{2}\int d^{5}x\sqrt{-g}~
          g^{MN}\partial_{M}\Phi\partial_{N}\Phi.
\end{eqnarray}
Using the metric (\ref{linee}), the equation of motion derived from
(\ref{action_scalar}) reads as
\begin{eqnarray}\label{EqOfScalar5D}
\frac{1}{\sqrt{-\hat{g}}}\partial_{\mu}(\sqrt{-\hat{g}}
   \hat{g}^{\mu \nu}\partial_{\nu} \Phi)
 +e^{-3A} \partial_{z}
  \left(e^{3A}\partial_z \Phi \right) = 0.
\end{eqnarray}
Then, by using of the KK decomposition
\begin{eqnarray}\label{ScaKKcom}
\Phi(x,z)=\sum_{n}\phi_{n}(x)\chi_{n}(z)e^{-3A/2},
\end{eqnarray}
the four-dimensional scalar fields $\phi_{n}(x)$ should satisfy the
four-dimensional massive Klein-Gordon equation:
\begin{eqnarray}
\label{4dKGEq}
\left(\frac{1}{\sqrt{-\hat{g}}}\partial_{\mu}(\sqrt{-\hat{g}}
   \hat{g}^{\mu \nu}\partial_{\nu}) -m_{n}^{2} \right)\phi_{n}(x)=0.
\end{eqnarray}
Therefore the equation for the scalar KK modes $\chi_{n}(z)$ can be
expressed as
\begin{eqnarray}
\left[-\partial^{2}_z+ V_{0}(z)\right]{\chi}_n(z)
  =m_{n}^{2} {\chi}_{n}(z),
  \label{SchEqScalar1}
\end{eqnarray}
which is a Schr\"{o}dinger equation with the effective potential
given by
\begin{eqnarray}
  V_0(z)=\frac{3}{2} A'' + \frac{9}{4}A'^{2}, \label{VScalar}
\end{eqnarray}
where $m_{n}$ is the mass of the KK excitation. It is clear that the
potential $V_{0}(z)$ defined in (\ref{VScalar}) is a
four-dimensional mass-independent potential.

The full five-dimensional action (\ref{action_scalar}) can be
reduced to one four-dimensional action for a massless scalar field
plus an infinite sum of massive scalar actions in four-dimension
\begin{eqnarray}
 S_{0}=- \frac{1}{2} \sum_{n}\int d^{4} x \sqrt{-\hat{g}}
     \bigg(\hat{g}^{\mu\nu}\partial_{\mu}\phi_{n}
           \partial_{\nu}\phi_{n}
           +m_{n}^2 \phi^{2}_{n}
     \bigg) \label{ScalarEffectiveAction}
\end{eqnarray}
when integrated over the extra dimension, in which it is required
that Eq. (\ref{SchEqScalar1}) is satisfied and the following
orthonormality conditions are obeyed:
\begin{eqnarray}
 \int^{+z_{b}}_{-z_{b}}
 \;\chi_m(z)\chi_n(z) dz=\delta_{mn}.
 \label{normalizationCondition1}
\end{eqnarray}

For the thick AdS brane, because the effective potential for the
scalar KK modes (\ref{VScalar}) is the same as the potential for the
gravity (\ref{VQM}), the localization of scalars is the same as the
situation of gravity. Hence, the four-dimensional massless scalar
(the zero mode) is not localized on the thick AdS brane for
$\delta>1$. And all the KK modes are massive bound states and are
localized on the brane. The mass spectrum of the KK modes is
discrete.

\subsection{Spin-1 vector fields}

\begin{figure}[htb]
\begin{center}
\includegraphics[width=7cm]{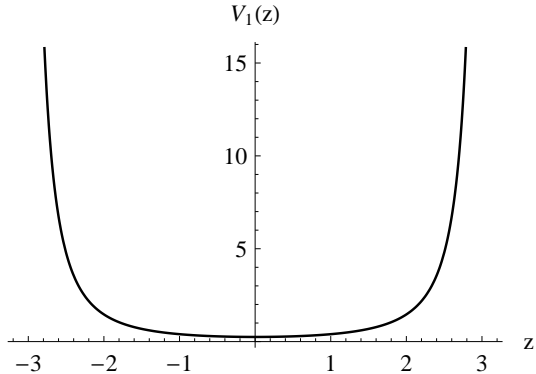}
\end{center}\vskip -5mm
\caption{The shape of the potential of the vector field $V_{1}(z)$.
 The parameters are set to $H=1$ and $\delta=2$. }
 \label{fig_Vec_V1Rho0}
\end{figure}

\begin{figure*}[htb]
\begin{center}
\subfigure[$n=1$]{\label{fig_Rho1c}
\includegraphics[width=3.5cm]{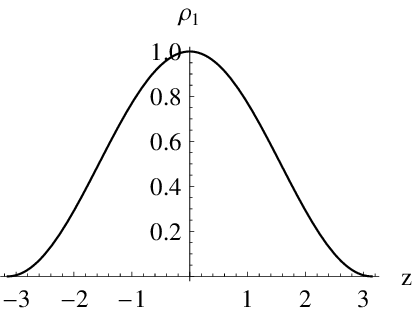}}
\subfigure[$n=2$]{\label{fig_Rho2c}
\includegraphics[width=3.5cm]{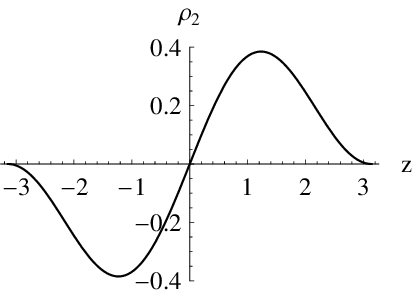}}
\subfigure[$n=3$]{\label{fig_Rho3c}
\includegraphics[width=3.5cm]{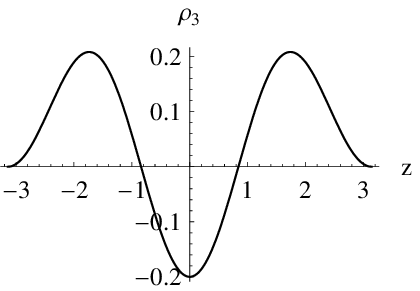}}
\subfigure[$n=4$]{\label{fig_Rho4c}
\includegraphics[width=3.5cm]{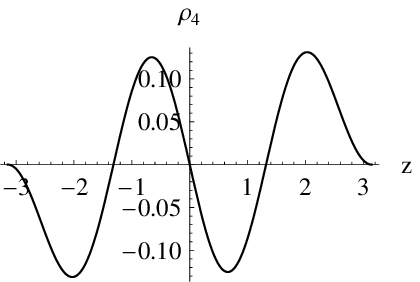}}
\end{center}\vskip -5mm
\caption{The shapes of the KK modes of the vector fields
$\rho_{n}(z)$. The parameters are set to $H=1$, and $\delta=2$.}
 \label{fig_rhonc}
\end{figure*}

We now turn to spin-1 vector fields. We begin with the
five-dimensional action of a vector field
\begin{eqnarray}\label{action_Vector}
S_{1} = -\frac{1}{4}\int d^{5}x \sqrt{-g}~ g^{M N}
 g^{RS}F_{MR}F_{NS},
\end{eqnarray}
where $F_{MN}=\partial_{M}A_{N}-\partial_{N}A_{M}$ is the field
tensor as usual. From this action the equations of motion are derived as
follows
\begin{eqnarray}
\frac{1}{\sqrt{-g}} \partial_{M} (\sqrt{-g} g^{M N} g^{R S} F_{NS})
= 0.
\end{eqnarray}
By using the background metric (\ref{linee}), the equations of
motion read as
\begin{eqnarray}
 \frac{1}{\sqrt{-\hat{g}}}\partial_\nu (\sqrt{-\hat{g}} ~
      \hat{g}^{\nu \rho}\hat{g}^{\mu\lambda}F_{\rho\lambda})
    +{\hat{g}^{\mu\lambda}}e^{-A}\partial_z
      \left(e^{A} F_{4\lambda}\right)  = 0, ~~\\
 \partial_\mu (\sqrt{-\hat{g}}~ \hat{g}^{\mu \nu} F_{\nu 4}) =
 0.~~
\end{eqnarray}
Because the fourth component $A_4$ has no zero mode in the effective
four-dimensional theory, we assume that it is $Z_2$-odd with respect
to the extra dimension $z$. Furthermore, in order to be consistent
with the gauge invariant equation $\oint dz A_4=0$, we choose
$A_4=0$ by using gauge freedom. Then, the action
(\ref{action_Vector}) can be reduced to
\begin{eqnarray}
S_1 = - \frac{1}{4} \int d^5 x \sqrt{-g} \Big[
        F_{\mu\nu}F^{\mu\nu}
        +2e^{-A} g^{\mu\nu} \partial_z A_{\mu} \partial_z A_{\nu}
       \Big].
\label{actionVector2}
\end{eqnarray}
With the decomposition of the vector field
\begin{eqnarray}\label{VecKKCom}
A_{\mu}(x,z)=\sum_n a^{(n)}_\mu(x)\rho_n(z)e^{-A/2}
\end{eqnarray}
and the orthonormality condition
\begin{eqnarray}
 \int^{+z_{b}}_{-z_{b}}  \;\rho_m(z)\rho_n(z)dz=\delta_{mn},
 \label{normalizationCondition2}
\end{eqnarray}
the action (\ref{actionVector2}) can be simplified as
\begin{eqnarray}
S_1 = \sum_{n}\int d^4 x \sqrt{-\hat{g}}~
       \bigg[\!\! &-& \!\! \frac{1}{4}\hat{g}^{\mu\alpha} \hat{g}^{\nu\beta}
             f^{(n)}_{\mu\nu}f^{(n)}_{\alpha\beta} \nonumber \\
      &-& \!\!\frac{1}{2}m_{n}^2 ~\hat{g}^{\mu\nu}
           a^{(n)}_{\mu}a^{(n)}_{\nu}
       \bigg],
\label{actionVector3}
\end{eqnarray}
where $f^{(n)}_{\mu\nu} = \partial_\mu a^{(n)}_\nu - \partial_\nu
a^{(n)}_\mu$ is the four-dimensional field strength tensor.
Therefore we have obtained a four-dimensional theory of a gauge
particle (massless) and infinite towers of massive vector fields. In
the above process, it has been required that the vector KK modes
$\rho_n(z)$ should satisfy the following Schr\"{o}dinger equation
\begin{eqnarray}
  \left[-\partial^2_z +V_1(z) \right]{\rho}_n(z)=m_n^{2}
  {\rho}_n(z),  \label{SchEqVector1}
\end{eqnarray}
with the mass-independent potential $V_1(z)$ given by
\begin{eqnarray}
\label{V_Vector}
 V_{1}(z)=\frac{H^{2}}{4\delta}\left[(2+\delta)\sec^{2}\bar{z}-\delta\right].
\end{eqnarray}
The potential also has a minimum $\frac{H^{2}}{2\delta}$ at $z=0$
and tends to positive infinity at $z=z_{b}$ for $\delta>1$. The
shapes of the potential $V_{1}(z)$ are plotted in
Fig.~\ref{fig_Vec_V1Rho0}. The potential $V_{1}(z)$ are always
positive, so the zero mode is not localized on the AdS brane.
However, the massive modes can be localized on the brane. Eq.
(\ref{SchEqVector1}) with this potential can be turned into the
following Schr\"{o}dinger equation:
\begin{eqnarray}\label{SchEqVector2}
\left[-\partial^2_z
    +\frac{H^{2}}{4\delta}(2+\delta)\sec^{2}\bar{z}
\right]\rho_{n}(z)=E_n^{2}\rho_{n}(z)
\end{eqnarray}
with $E^{2}_{n}=m_{n}^{2}+\frac{H^{2}}{4}$. The solutions of Eq.
(\ref{SchEqVector2}) for $\delta>1$ can be expressed as follows
\begin{eqnarray}\label{VectorN}
\rho_{n}\propto \cos^{1+\frac{\delta}{2}}(\bar{z})~
                   _{2}\text{F}_{1}\left[1-n, 1+n+\delta,
          \frac{3+\delta}{2}, \frac{1-\sin\bar{z}}{2}\right],
        ~~
\end{eqnarray}
where $n=1,2,\cdots~$,
and the discrete mass spectrum can be written as
\begin{eqnarray}
m_{n}=\frac{H}{\delta}\sqrt{n(n+\delta)},
\end{eqnarray}
so all the KK modes are bound states and can also be trapped on the
brane. We can find that the ground state is a massive state with
$m_{1}=\frac{H}{\delta}\sqrt{1+\delta}$. The shapes of the KK modes
are plotted in Fig.~\ref{fig_rhonc}.

\subsection{Spin-1/2 fermion fields}

Finally, we will study the localization of fermions on the thick AdS
brane. In five-dimensional space-time, fermions are four component
spinors and their Dirac structure can be described by the curved
space gamma matrices $\Gamma^{M}=e^{-A}(\gamma^{\mu},\gamma^{5})$,
where $\gamma^{\mu}$ and $\gamma^{5}$ are the usual flat gamma
matrices in the four-dimensional Dirac representation. The Dirac
action of a massless spin-1/2 fermion coupled to the background
scalar $\phi$ (\ref{scalarsolution}) is
\begin{eqnarray}
S_{1/2} = \int d^5 x \sqrt{-g} \Big[\bar{\Psi} \Gamma^M
          (\partial_M+\omega_M) \Psi
          -\eta \bar{\Psi} F(\phi)\Psi\Big],
\label{DiracAction}
\end{eqnarray}
where $\eta$ is a coupling constant. The nonvanishing components of
the spin connection $\omega_M$ for the background metric
(\ref{linee}) are
\begin{eqnarray}
  \omega_\mu =\frac{1}{2}A' \gamma_\mu \gamma_5
             +\hat{\omega}_\mu, \label{spinConnection}
\end{eqnarray}
with $\hat{\omega}_\mu$ the spin connection derived from the metric
$\hat{g}_{\mu\nu}(x)$. Then the equation of motion is given by
\begin{eqnarray}
 \Big[ \gamma^{\mu}(\partial_{\mu}+\hat{\omega}_\mu)
         + \gamma^5 \left(\partial_z  +2 A'\right)
         -\eta\; \text{e}^A F(\phi)
 \Big ] \Psi =0, \label{DiracEq1}
\end{eqnarray}
where $\gamma^{\mu}(\partial_{\mu}+\hat{\omega}_\mu)$ is the Dirac
operator on the AdS brane.

Now we would like to investigate the localization of the Dirac
spinor on the AdS brane by studying the above five-dimensional Dirac
equation. Because of the Dirac structure of the fifth gamma matrix
$\gamma^{5}$, we expect that the left- and right-chiral projections
of the four-dimensional part have different behaviors. From the
equation of motion (\ref{DiracEq1}), we will search for the
solutions of the general chiral decomposition
\begin{equation}\label{FerKKdecom}
 \Psi = \sum_n\Big[\psi_{L,n}(x) L_{n}(z)
 +\psi_{R,n}(x) R_{n}(z)\Big]\text{e}^{-2A},
\end{equation}
where $\psi_{L}=\frac{1-\gamma^5}{2}\psi$ and
$\psi_{R}=\frac{1+\gamma^5}{2} \psi$ are the left- and
right-chiral components of a four-dimensional Dirac field $\psi$, respectively. By
demanding $\psi_{L,R}$ satisfy the four-dimensional massive Dirac equations
$\gamma^{\mu}(\partial_{\mu}+\hat{\omega}_\mu)\psi_{L,R}
=m\psi_{R,L}$, we obtain the following coupled equations for the fermion KK modes $L_n$ and $R_n$:
\begin{subequations}\label{CoupleEq1}
\begin{eqnarray}
 \left[\partial_z + \eta\;\text{e}^A F(\phi) \right]L_{n}(z)
  &=&  ~~m R_{n}(z), \label{CoupleEq1a}  \\
 \left[\partial_z- \eta\;\text{e}^A F(\phi) \right] R_{n}(z)
  &=&  - m L_{n}(z). \label{CoupleEq1b}
\end{eqnarray}
\end{subequations}
From the above equations, we can obtain the Schr\"{o}dinger-like
equations for the KK modes of the left- and right-chiral fermions:
\begin{subequations}\label{SchEqFermion}
\begin{eqnarray}
  \big(-\partial^2_z + V_L(z) \big)L_{n}
            &=&m^2 L_{n},~~
   \label{SchEqLeftFermion}  \\
  \big(-\partial^2_z + V_R(z) \big)R_{n}
            &=&m^2 R_{n}.
   \label{SchEqRightFermion}
\end{eqnarray}
\end{subequations}
where the effective potentials of Eq. (\ref{SchEqFermion}) are given
by
\begin{subequations}\label{Vfermion}
\begin{eqnarray}
  V_L(z)&=& \big(\eta \text{e}^{A} F(\phi)\big)^2
     -\partial_z \big(\eta \text{e}^{A} F(\phi)\big), \label{VL}\\
  V_R(z)&=&   V_L(z)|_{\eta \rightarrow -\eta}. \label{VR}
\end{eqnarray}
\end{subequations}

For the purpose of getting the standard four-dimensional action for
a massless fermion and an infinite sum of the massive fermions,
\begin{eqnarray}
 S_{\frac{1}{2}} &=& \int d^5 x \sqrt{-g} ~\bar{\Psi}
     \Big[  \Gamma^M (\partial_M+\omega_M)
     -\eta F(\phi)\Big] \Psi  \nn \\
  &=&\sum_{n}\int d^4 x \sqrt{-\hat{g}}
    ~\bar{\psi}_{n}
      \Big[\gamma^{\mu}(\partial_{\mu}+\hat{\omega}_\mu)
        -m_{n}\Big]\psi_{n},~~~
\end{eqnarray}
we need the following orthonormality conditions for $L_n$ and $R_n$:
\begin{eqnarray}
 \int_{-z_{b}}^{z_{b}} L_m L_ndz
   &=& \delta_{mn}, \label{orthonormalityFermionL} \\
 \int_{-z_{b}}^{z_{b}} R_m R_ndz
   &=& \delta_{mn}, \label{orthonormalityFermionR}\\
 \int_{-z_{b}}^{z_{b}} L_m R_ndz
   &=& 0. \label{orthonormalityFermionR}
\end{eqnarray}

\begin{figure*}[htb]
\begin{center}
\subfigure[$V_{L}(z)$]{\label{fig_VL_3}
\includegraphics[width=7cm]{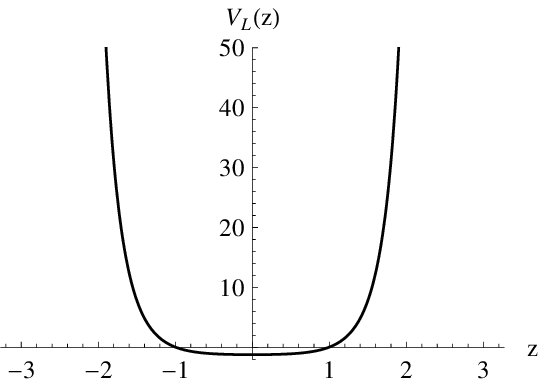}}
\subfigure[$V_{R}(z)$]{\label{fig_VR_3}
\includegraphics[width=7cm]{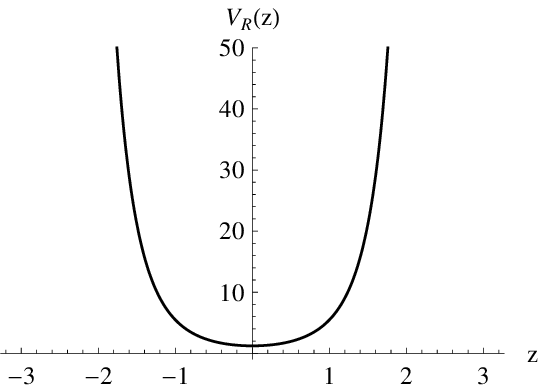}}
\end{center}\vskip -5mm
\caption{The shapes of the potentials of the left- and right-chiral
fermion KK modes. The parameters are set to $\delta=2$, $H=1$ and
$\eta=1$.}
 \label{fig_VLR}
\end{figure*}

From Eqs. (\ref{SchEqFermion})and (\ref{Vfermion}), we can see that,
in order to localize the left- and right-chiral fermions, there must
be some kind of scalar-fermion coupling. This situation is similar
to the one in Refs. \cite{BajcPLB2000,Oda_PRD}, in which the authors
introduced the mass term $m\epsilon(z)\bar{\Psi}\Psi$ for the
localization of the fermion fields on a brane. For the thick branes
arising from a real scalar field $\phi$, the scalar-fermion coupling
such as $\eta\bar{\Psi}\phi\Psi$, $\eta\bar{\Psi}\phi^{k}\Psi$,
$\eta\bar{\Psi}\tan^{1/s}(\phi)\Psi$, etc., can be introduced for
fermion localization
\cite{Volkas0705.1584,Ringeval,Liu0708,Liu0907.0910}. For the thick
brane generated by two scalar $\phi$ and $\chi$, in order to
localize the fermions, the coupling terms
$\eta\bar{\Psi}\phi\chi\Psi$ and
$\eta\bar{\Psi}\phi\Psi+\eta'\bar{\Psi}\chi\Psi$ were introduced in
Ref. \cite{zhao_CQG}. Moreover, if we demand that $V_{L}(z)$ and
$V_{R}(z)$ are $Z_{2}$-even with respect to the extra dimension $z$,
$F(\phi)$ should be an odd function of the ``kink'' $\phi(z)$. In
this paper, we choose the simplest Yukawa coupling: $F(\phi)=\phi$.
Then the explicit forms of the potentials (\ref{Vfermion}) are
\begin{eqnarray}
\label{VL}
V_{L}(z)&=& \eta^{2}\phi_{0}^{2}\cos^{-2\delta}(\bar{z})
           \text{arcsinh}^{2}(\tan\bar{z})\nonumber\\
         &~& - \frac{\eta H\phi_{0}}{\delta \cos^{1+\delta}(\bar{z})}
           \Big[\delta\sin\bar{z} ~\text{arcsinh} (\tan\bar{z})+1 \Big],~~~\\
V_{R}(z)&=& V_{L}(z)|_{\eta\rightarrow -\eta}~.
 \label{VR}
\end{eqnarray}
At the location of the AdS brane, both of the potentials for the
left- and right-chiral fermions have minimum $V_{L}(z=0)=-{\eta
H\phi_{0}}/{\delta}$ and $V_{R}(z=0)={\eta H\phi_{0}}/{\delta}$, and
the potentials $V_{L,R}(z\rightarrow\pm z_{b})\rightarrow +\infty$
as shown in Fig.~\ref{fig_VLR}.


\begin{figure*}[htb]
\begin{center}
\subfigure[$I(z)$]{\label{fig_Iz}
\includegraphics[width=7cm]{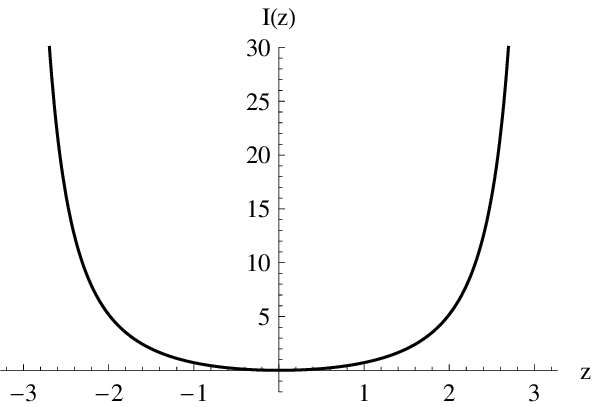}}
\subfigure[$L_{0}$]{\label{fig_L0_3}
\includegraphics[width=7cm]{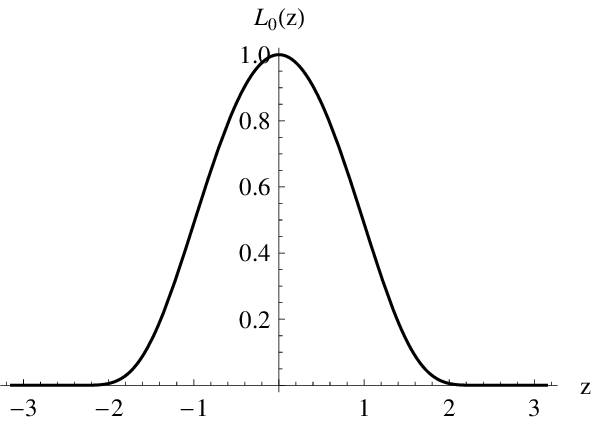}}
\end{center}\vskip -5mm
\caption{The shape of the exponential factor $I(z)$ and the
left-chiral fermion zero mode $L_{0}$. The parameters are set to
$H=1$, $\eta=1$ and $\delta=2$.}
 \label{fig_L0}
\end{figure*}

From Fig.~\ref{fig_VLR}, only the potential of the left-chiral
fermions has negative value at the location of the thick brane for
the positive coupling constant $\eta$, so only the left-chiral
fermion zero mode can be localized on the brane. By setting $m=0$ in
Eq. (\ref{CoupleEq1}), the left-chiral fermion zero mode can be
obtained:
\begin{eqnarray}
 \label{L0}
 L_{0}(z)&\propto& \exp[-\eta I(z)],
\end{eqnarray}
where the exponential factor $I(z)$ can be expressed as
\begin{eqnarray*}
I(z)&=&\int^{z}_{0} dz'\text{e}^{A(z')}\phi(z')\nonumber\\
    &=&\frac{2^{1-\delta}\phi_{0}\delta}{H(\delta-1)^{2}}
    \left\{ _{3}\text{F}_{2} \Big( 1-\delta, \frac{1-\delta}{2}, \frac{1-\delta}{2};
        \frac{3-\delta}{2}, \frac{3-\delta}{2}; -1\Big)\right.\nonumber\\
     ~~~&& - \varrho^{1-\delta}(z)\left[ _{3}\text{F}_{2} \Big( 1-\delta, \frac{1-\delta}{2}, \frac{1-\delta}{2};
        \frac{3-\delta}{2}, \frac{3-\delta}{2}; -\varrho^{2}(z)\Big)
        \right. \nonumber\\
     ~~~&& + \left.\left.
        (\delta-1)~_{2}\text{F}_{1}
         \Big( \frac{1-\delta}{2}, 1-\delta, \frac{3-\delta}{2},
         -\varrho^{2}(z)\Big)\ln\Big(\varrho(z)\Big)
         \right]\right\}
\end{eqnarray*}
with $\varrho(z)=\sec\bar{z}+\tan\bar{z}$.
The curve for the exponential factor is plotted in
Fig.~\ref{fig_Iz}. The exponential factor $I(z)$ tends to positive
infinity, so the left-chiral fermion zero mode $L_{0}\rightarrow 0$
as $z\rightarrow \pm z_{b}$, and is localized on the AdS brane,
which is shown in Fig.~\ref{fig_L0_3}.

\begin{figure*}[htb]
\begin{center}
\subfigure[$n=0$]{\label{fig_L0}
\includegraphics[width=3.5cm]{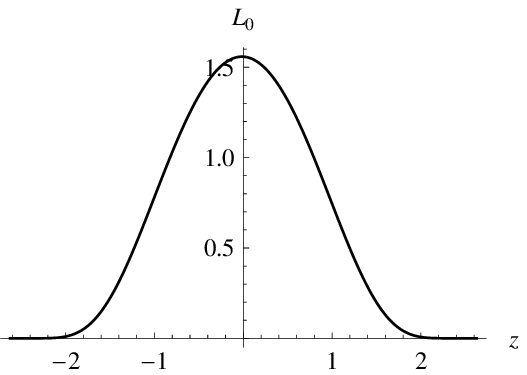}}
\subfigure[$n=1$]{\label{fig_L1}
\includegraphics[width=3.5cm]{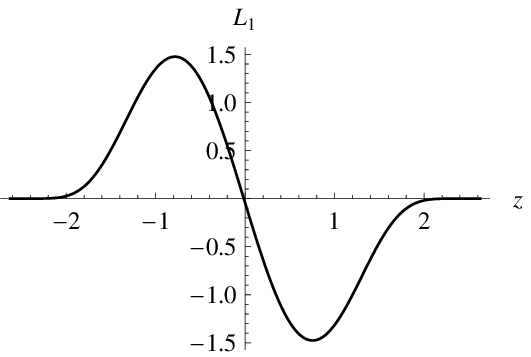}}
\subfigure[$n=2$]{\label{fig_L2}
\includegraphics[width=3.5cm]{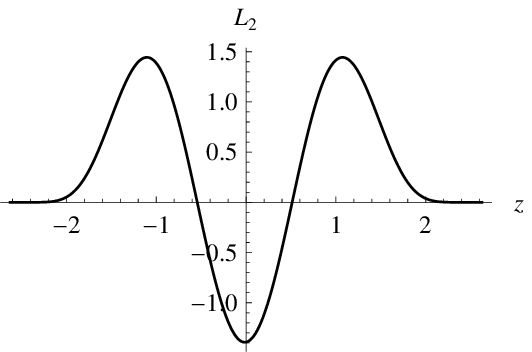}}
\subfigure[$n=3$]{\label{fig_L3}
\includegraphics[width=3.5cm]{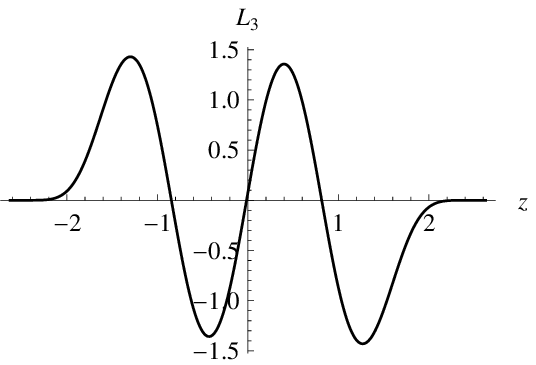}}
\end{center}\vskip -5mm
\caption{The shapes of the KK modes of the left-chiral fermions. The
parameters are set to $H=1$, $\eta=1$ and $\delta=2$.}
 \label{fig_Ln}
\end{figure*}

\begin{figure*}[htb]
\begin{center}
\subfigure[$n=1$]{\label{fig_R1}
\includegraphics[width=3.5cm]{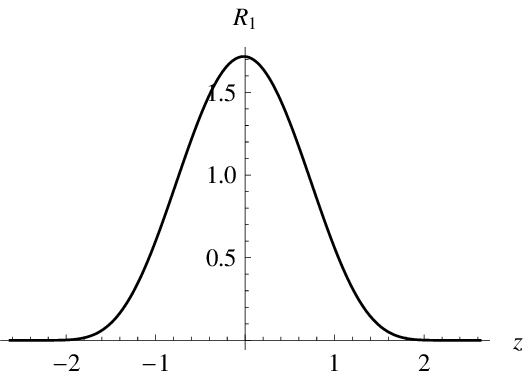}}
\subfigure[$n=2$]{\label{fig_R2}
\includegraphics[width=3.5cm]{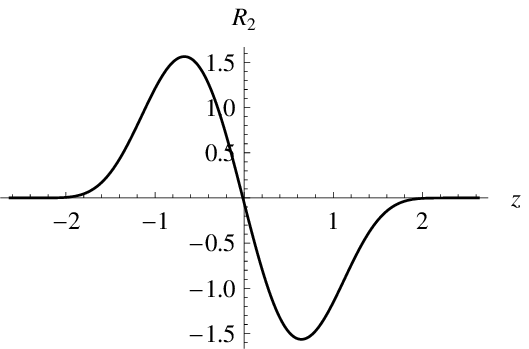}}
\subfigure[$n=3$]{\label{fig_R3}
\includegraphics[width=3.5cm]{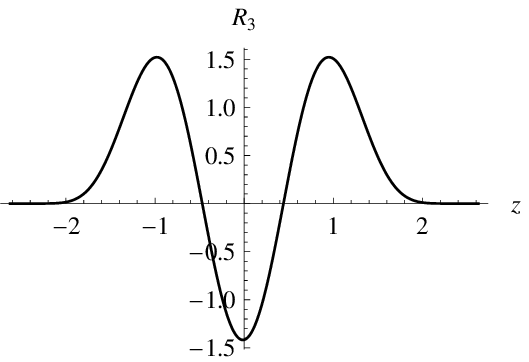}}
\subfigure[$n=4$]{\label{fig_R4}
\includegraphics[width=3.5cm]{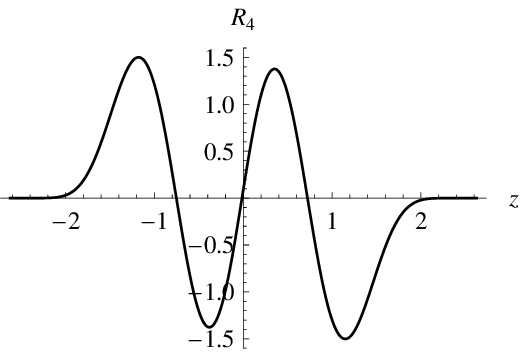}}
\end{center}\vskip -5mm
\caption{The shapes of the KK modes of the right-chiral fermions.
The parameters are set to $H=1$, $\eta=1$ and $\delta=2$.}
 \label{fig_Rn}
\end{figure*}

\begin{figure}[htb]
\begin{center}
\includegraphics[width=8cm]{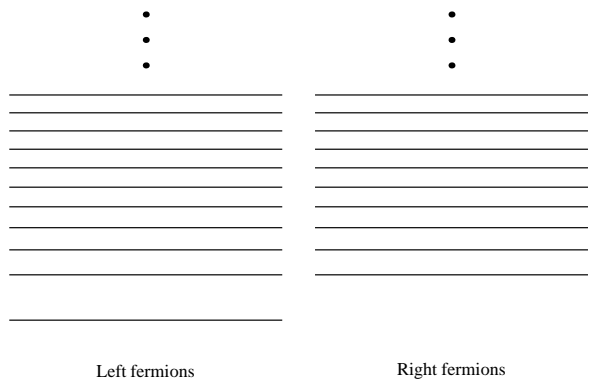}
\end{center}\vskip -5mm
\caption{The $m_{L_{n},R_{n}}$ spectra of the left- and right-chiral
fermions. The parameters are set to $H=1$, $\eta=1$ and $\delta=2$.}
 \label{fig_FermionMn}
\end{figure}

For the massive KK modes of the left- and right-chiral fermions,
Eqs. (\ref{SchEqLeftFermion}) and (\ref{SchEqRightFermion}) cannot
be analytically solved. But we can solve them by numerical method.
The KK modes of the left- and right-chiral fermions are plotted in
Figs.~\ref{fig_Ln} and \ref{fig_Rn}, respectively. All the KK modes
for both the left- and right-chiral fermions are bound states. The
discrete mass spectra for the KK modes  are calculated as follows:
\begin{subequations}\label{FermionMass}
\begin{eqnarray}
 \label{FermionMass_a}
 m_{L_{n}}&=& \{0, 1.78, 2.76, 3.63, 4.45, 5.22, 5.98, 6.71,\nonumber\\
          &&~~~~~     7.43, 8.14, 8.84,\cdots\}, \\
 \label{FermionMass_b}
 m_{R_{n}}&=& \{~~~ 1.78, 2.76, 3.63, 4.45, 5.22, 5.98, 6.71,\nonumber\\
           &&~~~~~ 7.43, 8.14, 8.84, \cdots \},
\end{eqnarray}
\end{subequations}
where the parameters are set to $H=1$, $\eta=1$ and $\delta=2$.
Hence the ground state of the left-chiral fermions is the massless
mode; however, the ground state of the right-chiral fermions is a
massive mode. The mass spectra are also shown in
Fig.~\ref{fig_FermionMn}. So we can obtain the four-dimensional
massless left-chiral fermion and the massive Dirac fermions
consisting of the pairs of the left- and right-chiral KK modes
coupling together through mass terms.


\section{Conclusion and discussion}
\label{SecConclusion}

In this paper, we first reviewed a thick AdS brane embedded in
five-dimensional space-time, in which the behavior of the warp
factor is related to a parameter $\delta$. For $\delta<0$, the warp
factor tends to zero at the boundaries of the extra dimension, while
the energy density has no lump at $z\approx0$, which cannot be
considered as a thick brane. For $\delta>1$, the warp factor tends
to infinity at the boundaries, and the energy density has a lump at
$z\approx0$, which indicates that the AdS thick brane locates at
$z\approx0$. Hence, we only consider the case of $\delta>1$. Then we
studied the mass-independent potentials of the KK modes for gravity
and various spin fields in the corresponding Schr\"{o}dinger
equations. In this way, the localization and mass spectra of gravity
and various matters with spin-0, 1 and 1/2 on this kind of AdS brane
were investigated.

For gravity, the potential of the KK modes in the corresponding
Schr\"{o}dinger equation is divergent when far away from the brane.
Such potential suggests that the mass spectrum of the gravity KK
modes consists of an infinite number of discrete bound states.
Although, the zero mode of gravity is not localized on the AdS
brane, the massive modes can be localized on the brane. The scalar
perturbations of the thick AdS brane has been analyzed, and as a
result, it is shown that the brane is stable under the scalar
perturbations.

For spin-0 scalars and spin-1 vectors, the character of localization
is similar to the case of gravity; i.e., all the scalar and vector
KK modes are bound states, and the zero modes are not localized on
the brane.

For spin-1/2 fermions, in order to localize the left- and
right-chiral fermions, we introduced the usual Yukawa coupling
$\eta\bar{\Psi}\phi\Psi$ with a positive coupling constant $\eta$.
Both potentials have the same asymptotic behavior:
$V_{L,R}(z\rightarrow\pm z_{b})\rightarrow +\infty$, and only the
potential of the left-chiral fermion KK modes has a finite negative
well at $z=0$. So only the left-chiral fermion zero mode could be
localized on the brane; i.e., there exists only the four-dimensional
massless left-chiral fermion. And both the left- and right-chiral
fermion KK modes have an infinite number of bound states. Since a
pair of left- and right-chiral KK modes couple together through a
mass term to become a four-dimensional Dirac fermion, a series of
four-dimensional Dirac fermions with a discrete mass spectrum could
be obtained on the AdS branes.

\section*{Acknowledgement}

We are grateful to the referee, whose comments led to the improvement of this paper.
This work was supported by the Program for New Century Excellent Talents in
University, the Huo Ying-Dong Education Foundation of Chinese Ministry of Education
(No. 121106), the National Natural Science Foundation of China (No. 11075065), the
Doctoral Program Foundation of Institutions of Higher Education of China (No.
20090211110028), the Key Project of Chinese Ministry of Education (No. 109153), and
the Natural Science Foundation of Gansu Province, China (No. 096RJZA055).

\end{document}